\documentclass[pre,twocolumn,a4paper,showpacs]{revtex4}
\usepackage{amssymb}
\usepackage{graphicx}
\usepackage{epsfig}
\usepackage{subfigure}
\begin{document}

\title{Revisiting the effect of external fields in Axelrod's model of social dynamics}

\author{Lucas R. Peres and Jos\'e F. Fontanari}
\affiliation{Instituto de F\'{\i}sica de S\~ao Carlos,
  Universidade de S\~ao Paulo,
  Caixa Postal 369, 13560-970 S\~ao Carlos, S\~ao Paulo, Brazil}


\begin{abstract}
The study of the effects of spatially uniform fields on the steady-state properties of Axelrod's model has 
yielded  plenty of controversial results. Here we re-examine the impact of this type of field for a selection 
of parameters  such that the field-free steady state of the model is heterogeneous or multicultural.  Analyses of
both  one and two-dimensional versions of Axelrod's model indicate that, contrary to previous claims in the literature,  the
steady state remains heterogeneous regardless of the value of the field strength. Turning on the field 
leads to a discontinuous decrease  on the number of cultural domains, which we argue is
due to the instability of zero-field heterogeneous absorbing configurations.  We find, however, 
that spatially nonuniform fields that implement  a consensus rule among the neighborhood of the agents
enforces  homogenization.  Although the overall  effects of the fields are essentially the same irrespective of the dimensionality of the model, 
we argue that the dimensionality has a significant impact on  the stability of the field-free  homogeneous steady state. 
\end{abstract}

\maketitle

\section{Introduction}

Axelrod's   model   of social dynamics was introduced to explore the mechanisms behind the persistence of
cultural differences in a society \cite{Axelrod_97}.  The agents  are represented by  strings  of
cultural features of length $F$, where each feature can adopt a certain number $q$ of distinct traits. The interaction between any two agents takes place
with probability proportional to their cultural similarity, i.e., proportional to the number of traits they have in common.
The analysis of this model by the statistical physics community has revealed a rich  dynamic behavior 
with a nonequilibrium  phase transition separating the culturally heterogeneous   from the culturally  homogeneous regime
  \cite{Castellano_00,Castellano_09,Barbosa_09}. 

An interesting characteristic of Axelrod's model, which sets it  apart from most
lattice models that exhibit nonequilibrium phase transitions \cite{Marro_99}, is  that all stationary 
states of the dynamics are absorbing states, i.e., the dynamics always freezes in one of these states  \cite{Castellano_00}. In fact,
according to the rules of Axelrod's model, two neighboring agents who do not have any cultural trait 
in common cannot interact and the interaction between agents who share all their cultural traits does not result in any change.
 Hence at the stationary state we can guarantee that any pair of neighbors are either identical 
or completely different  regarding their cultural features. This allows us to  easily identify the stationary regime, which is a major
problem in the characterization of nonequilibrium phase transitions  \cite{Marro_99}. The problem, however,  is that the
dynamics can take a very large time to freeze to a homogeneous configuration for some
 initial conditions \cite{Castellano_00, Vazquez_07}, which  is the  main reason there are so few numerical estimates of the transition lines of
 the phase diagram of Axelrod's model \cite{Barbosa_09}.
 
 From the perspective of the statistical physics, the appealing feature of Axelrod's model in a lattice of dimension $d$ is the existence of a threshold value $q_c = q_c \left ( F \right )$ with
 $F > 1$  below which the stationary regime is monocultural (i.e., spatially homogeneous)  and above which  multicultural
(i.e., spatially heterogeneous). 
This result holds true for both the two-dimensional \cite{Castellano_00,Barbosa_09} and the
 one-dimensional \cite{Vilone_02} versions of Axelrod's model. We recall that the sources of
disorder in this model are the stochastic update sequence and the choice of the initial configuration: 
it is the competition between the disorder of the initial configuration and  the ordering bias of the local  interactions that 
is responsible for the nontrivial threshold phenomenon.

The  introduction  of an external  global field  to influence the agents' beliefs aiming at modeling the 
effect of the mass media  \cite{Shibanai_01}  resulted in a surprisingly difficult and controversial problem  since
the external field favored the heterogeneous instead of the homogeneous  regime  as one  would naively   expect (see \cite{Peres_11} for
an explanation of this finding). In addition, a considerable amount  of effort has been devoted to  searching for a threshold on the intensity of the media influence such that above
that threshold, the community  is multicultural and below it, the community is mono-cultural  (see, e.g., \cite{Avella_05,Avella_06,Mazzitello_07,Candia_08}). 
We have shown, however, that this threshold is 
an artifact of finite lattices, and that even a vanishingly small media influence is sufficient to produce   cultural
diversity    in a region of the parameters space  where  the homogeneous regime is dominant  in the absence of the media \cite{Peres_11,Peres_10a}.

In this paper we address another curious finding regarding the effect of the media (or external field) in Axelrod's model:
if the control parameters are such that the stationary regime is multicultural at zero field, then turning  the
field on  will lead to a  homogeneous state in the limit of vanishingly small  field intensity \cite{Avella_06}. Here we argue
that this claim is not correct and that the multicultural regime remains  multicultural, though with a reduced  cultural diversity,
regardless of the intensity of the global external field.  By global field we mean a field that is spatially uniform (i.e., it is the same for all agents) although 
not necessarily time independent  \cite{Shibanai_01}.

Over and above the re-examination of far-out claims on the effect of external fields in Axelrod's model, in this contribution we
show that the effect of the  media in the one-dimensional model is  qualitatively identical to the two-dimensional model, 
which has been the sole focus of systematic analyses up to now (see  \cite{Vilone_02} for the study of the media-free one-dimensional
Axelrod's model). Since simulations of the one-dimensional model are fast we will use their results as 
clues to expose the properties of the stationary state of the two-dimensional model in the computationally prohibitive regime of  large lattices.

The paper is organized as follows. In Section \ref{sec:model} we briefly present the basic elements of Axelrod's model and  describe the different
types of external global fields (media) we study in this  paper. In Section \ref{sec:1d} we show the results of the simulations for the one-dimensional model in the two
cases  of interest, namely, when  the media-free absorbing configurations are homogeneous and when they are heterogeneous. In the first case, we show that, similarly to the 
two-dimensional version \cite{Peres_11,Peres_10a}, even a vanishingly small field intensity is sufficient to break up the homogeneous steady state. 
In the second case we show that the external field reduces the  number of cultural domains but the steady state remains heterogeneous, regardless of
the field strength. This conclusion is corroborated by the simulations of the two-dimensional model described in Section \ref{sec:2d}. A
different type of field -- a spatially non-uniform field that implements the consensus or majority  rule among the neighborhood of the agents  --  is discussed in 
a separate section because it cannot be interpreted as a media and produces results completely distinct from the spatially uniform fields (see Section \ref{sec:disc}).
Finally, in Section \ref{sec:conc} we summarize our main findings and present our concluding remarks.

\section{Model}\label{sec:model}

In the original formulation of Axelrod's model \cite{Axelrod_97}, which we will adhere to here, each agent is characterized by a set of $F$ cultural features which can take on
$q$ distinct values. In the two-dimensional version, the agents are fixed in the sites of a square lattice of linear size $L$ with free boundary conditions
(i.e., agents in the corners of the lattice interact with two neighbors, agents in the sides  with three, 
and agents in the bulk with four nearest neighbors), whereas in the one-dimensional variant the agents are fixed in the sites of a chain of length $L$ with
the same boundary conditions.
From our perspective, the advantage of using free boundary conditions is the easy of implementing 
Hoshen and Kopelman algorithm for counting the number of clusters in a lattice \cite{Stauffer_92},  but since we are interested in the
properties of the steady state for very large lattice sizes, the choice of the boundary conditions is largely irrelevant for Axelrod's model.

The initial configuration is completely random with the features of each agent given by  
random integers drawn uniformly between $1$ and $q$. At each time we pick an agent at random -- the target agent -- as well as one of its neighbors. These two 
agents interact with probability equal to  their cultural similarity, defined as the fraction of 
common cultural features. An interaction consists of selecting at random one of the distinct features, and making the
selected feature of the target agent equal  to its neighbor's corresponding trait. This procedure is repeated until 
the system is frozen into an absorbing configuration.  Thus Axelrod's model can be viewed as  $F$ coupled voter models
\cite{Liggett_85}.

The introduction of an external field  or global media in the standard model follows the ingenious suggestion 
 of adding   a virtual  agent which interacts with all agents in the lattice and whose cultural traits reflect the media 
message \cite{Shibanai_01}. In the original version,  each 
cultural feature of the virtual agent has the trait which is the most common in the population --
the consensus opinion. Henceforth we will refer to this type of external field as the consensus field.
The second type of field we consider is constant in time, i.e., the media message is
fixed from the outset, so it really models some alien influence impinging on the population. We will refer to this field as
the static field. Explicitly, we generate the culture vector of the virtual agent at random and keep it fixed
during the dynamics \cite{Avella_05,Avella_06}. (These two types of field were referred to as global and external media by 
Ref.\ \cite{Avella_06}, but since both fields are global and external here we opt to use a more informative nomenclature.)

Regardless of the type of external field, the interaction of the media (virtual agent) with the real agents is governed by
the control parameter
$p \in \left [ 0,1 \right ] $, which may be interpreted as a measure of the strength of the external field influence. 
As in the original Axelrod's
model, we begin by choosing  a target agent at random, but now it  can interact with the media with probability $p$ 
or with its neighbors with probability $1-p$.  Since we have defined the media as a virtual  agent, 
the interaction follows exactly the same rules as before.  The  media-free model is recovered by setting $p=0$.

Because of the unusually large times needed for some initial configurations to relax to a homogeneous absorbing configuration -- relaxation to heterogeneous
configurations is typically very fast -- the simulation of the dynamics of  Axelrod's model must be made as efficient as possible. 
In our simulations we consider two lists  of active bounds. The first list (list $A$)  is composed by the active bounds that connect real agents, whereas
the second list (list $B$) contains the active bounds that connect the virtual agent (media) with the real ones. In both cases, an active bound is  defined as a bound 
that connects two agents that have at least one feature in common  and at least one feature distinct from each other. 
Here instead of picking the 
target agent at random we first select one of  the two lists -- list $A$ with probability $1-p$ and list $B$ with probability $p$ -- and then pick a bound at random from the 
selected list. In case of a bound from list $B$, the target agent is of course the real one but if the bound belongs to list $A$ we choose the target agent at random from the two
options. In the case that
the cultural features of the target agent are modified by the interaction with its neighbor, we 
need to re-examine the active/inactive status of all bounds associated to the target agent  so as
to update the lists of active bounds.  The dynamics is frozen when the two lists of active bounds are emptied.

\begin{figure}[!ht]
  \begin{center}
\subfigure{\includegraphics[width=0.48\textwidth]{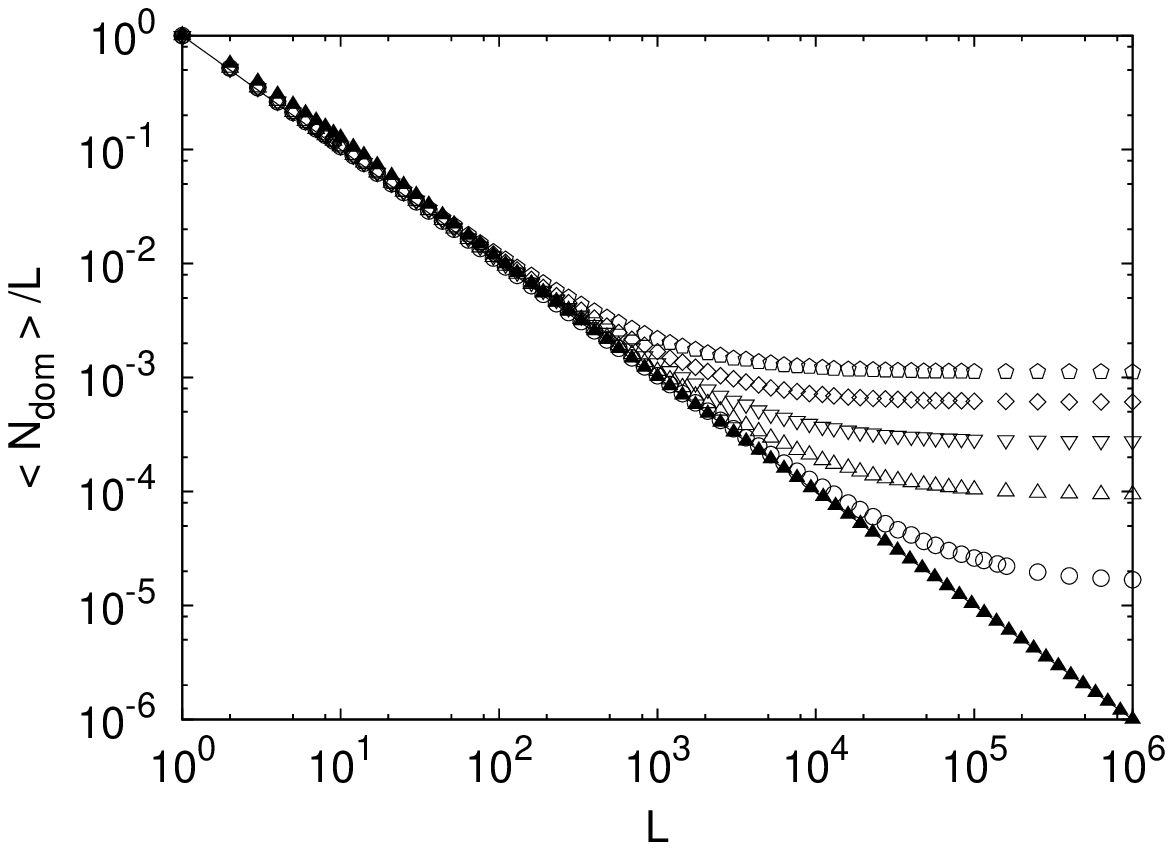}}
\subfigure{\includegraphics[width=0.48\textwidth]{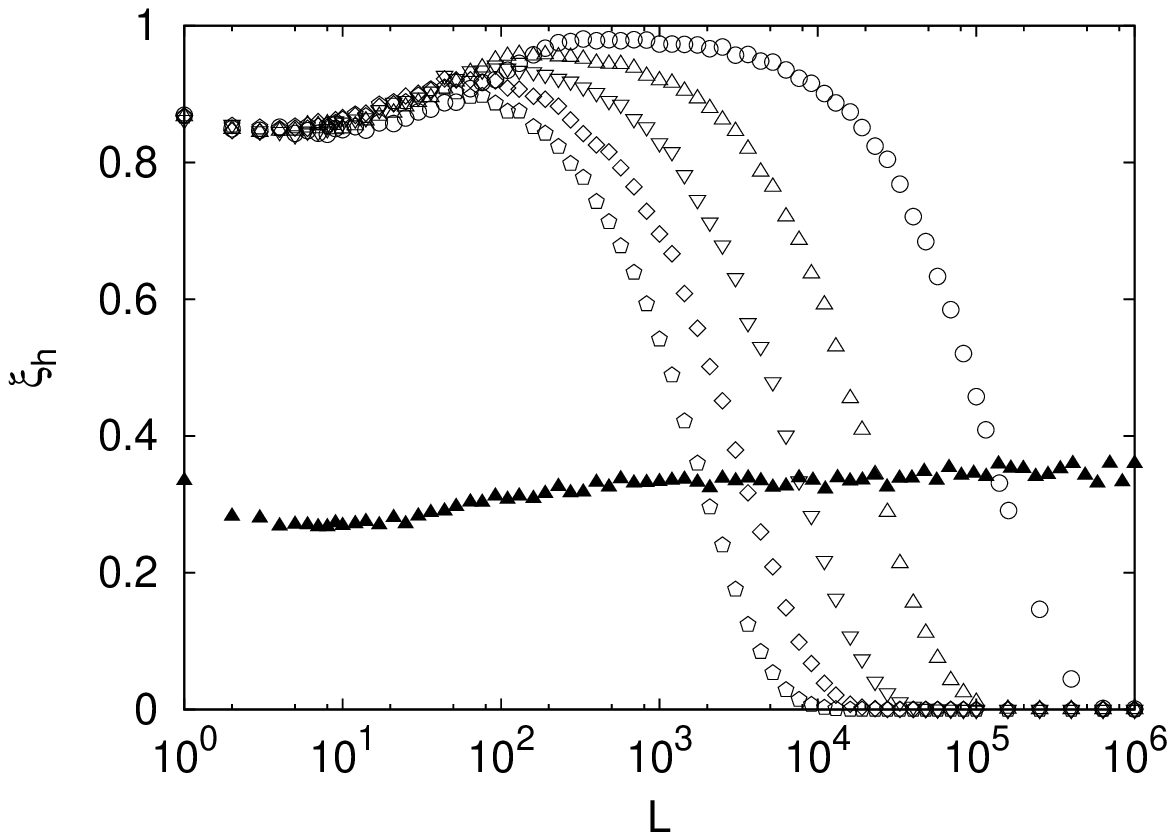}}
\subfigure{\includegraphics[width=0.48\textwidth]{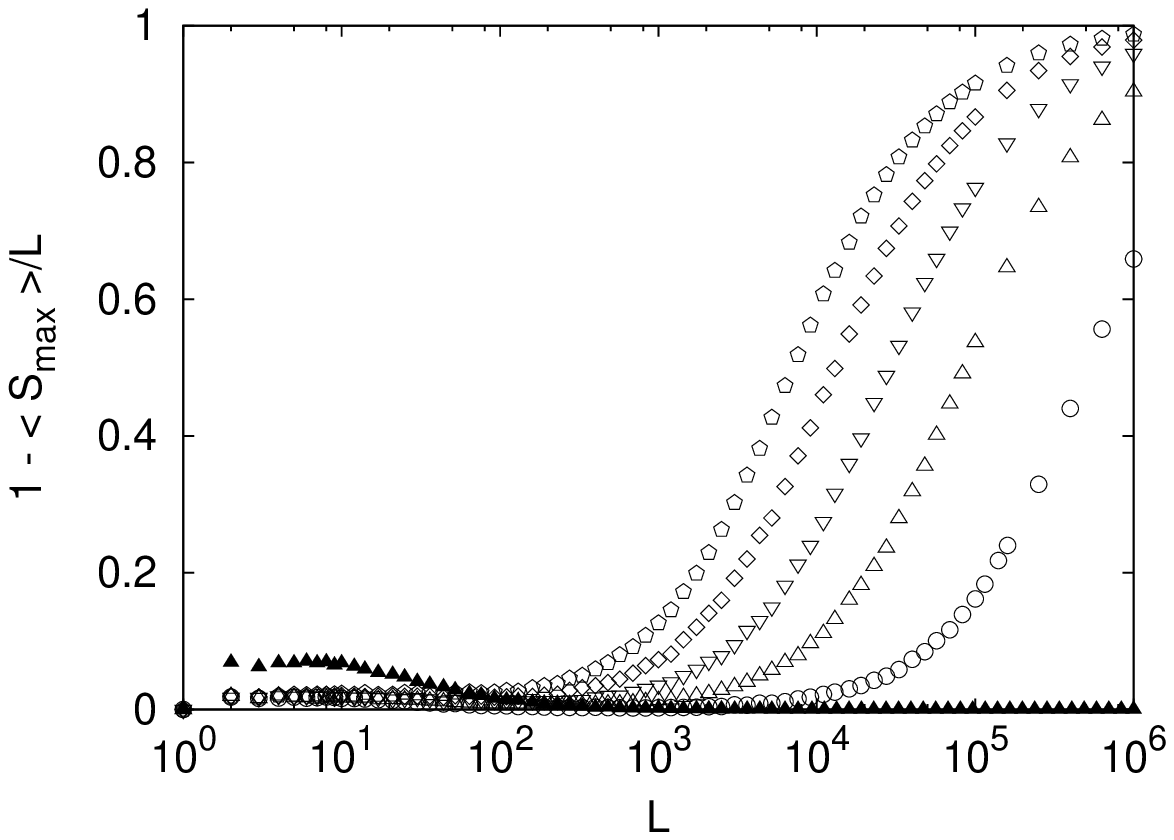}}
  \end{center}
\caption{Results for the static media in the one-dimensional model with $F = 5$ and $q = 3$ and for media
strengths (unfilled symbols, bottom to top in the upper and lower panels and  top to bottom in the middle panel) $p = 0.01, 0.02, 0.03, 0.04$ and $0.05$. 
The filled triangles are the  media-free  ($p=0$) results.
Here $\langle N_{dom} \rangle $ is  the average  number of domains,
$\xi_h$ is the fraction of runs that ended up in homogeneous absorbing configurations, and
 $\langle S_{max} \rangle $ is the average size of the largest domain.
The straight line in the upper panel is $1/L$. }
\label{fig:1}
\end{figure}

\section{External fields in the  one-dimensional model}\label{sec:1d}

In the attempt to make Axelrod's model more  `realistic', researchers  have studied the model in a variety of complex networks
(see, e.g., \cite{Greig_02,Klemm_03a,Avella_10}), with the usual result that the multicultural regime is destabilized by the increase of the connectivity of
the network and so the stationary regime is homogeneous regardless of the values of the parameters $q$ and $F$.  It is curious that   the simple 
one-dimensional variant that preserves the phase transition \cite{Vilone_02}  received comparatively  almost no  attention. In this section we show that the effect of
external fields is essentially the same in one and two dimensions as far as the favoring of the homogeneous or heterogeneous is concerned. The dimensionality 
introduces some distinctive effects, however, as we discuss next.

\subsection{Field-free homogeneous regime}

We begin our analysis with the once puzzling   situation in which the homogeneous media-free stationary regime becomes
heterogeneous under the influence of the external media  \cite{Shibanai_01}. Since this problem was extensively studied in the
two-dimensional lattice \cite{Avella_05,Avella_06,Peres_11,Peres_10a} we present here only  a brief analysis of the effect of the static media, 
aiming at highlighting the similarities and  differences  between the one and two-dimensional results.

To characterize the steady state of Axelrod's model we focus on two basic statistical measures, namely,
 the normalized average number of domains $N_{dom}/L$ and the average relative size of the  domains that do not belong to the largest domain $1 - S_{max}/L$. These quantities are shown 
 in Fig.\ \ref{fig:1} together with the fraction of runs  trapped into homogeneous absorbing configuration $\xi_h$.  A domain or cluster is a bounded region 
in which all agents share the same culture.
We note that in this figure,  as well as  in the next figures of this paper, the statistical error bars are smaller or at most equal to the symbol sizes.
Typically, each symbol represents the result of the average over $10^3$ to $10^4$ independent runs of the stochastic  dynamics. 

The first point to note is the remarkable similarity between the results presented  in the upper panel of  Fig.\ \ref{fig:1} for the one-dimensional and those
for the two-dimensional model  \cite{Peres_11}. For a fixed  finite value of $L$ there seems to exist  a threshold value  for the media
strength $p= p_c$ below which the regime is mono-cultural  \cite{Avella_05,Avella_06}. The quantity $\xi_h$ shown in the middle panel  illustrates this
somewhat odd predominance of the homogeneous  absorbing configurations, in which the agents are identical to the static media, for lattices of intermediate size
and small $p$ (the case $p=0$ is discussed below).
However,  as illustrated in the figure this `threshold'  decreases with increasing $L$ and so it is a finite size effect. 
Thus our conclusion is that in the  thermodynamic limit even a vanishingly small field is sufficient to break  up the mono-cultural regime. More pointedly,
extrapolating  the ratio $\langle N_{dom} \rangle /L $ (upper panel of Fig.\ \ref{fig:1}) for $L \to \infty$ and plotting the result against $p$  in a log-log graph yields
$\lim_{L \to \infty} \langle N_{dom} \rangle /L  \sim p^{2.7} $ in the limit  $p \to  0$. 

The middle and  lower panels of Fig.\ \ref{fig:1} reveal some remarkable differences between the one and two-dimensional models.  Firstly,
for $p=0$ only about 35\% of the samples (random initial configurations) ended up into  strictly homogeneous absorbing configurations, whereas
for the  two-dimensional model this happens for all samples in the homogeneous regime. The reason  we keep referring  to this regime as the
homogeneous  regime is that the order parameters $N_{dom}/L $ and $S_{max}/L$ take on values compatible with a homogeneous phase. 
This only happens because  the heterogeneous absorbing configurations are composed of a single macroscopic domain together with a non-extensive
number of microscopic domains. Hence the probability that a randomly chosen site in such configuration belongs to the largest domain is $1$
in the thermodynamic limit.
Secondly and more importantly,  in the two-dimensional model
the  fraction of the sites that are not part of the largest domain (the media in that case) is finite  and approaches zero for $p \to 0$ \cite{Peres_11}, 
whereas in the one-dimensional model
that fraction tends to unity regardless of  the field strength $p>0$.  This means that the effect of the  field 
 in one-dimension is quite extreme -- a vanishingly small field  is enough to completely destroy
the uniform regime, producing a fragmented configuration composed of microscopic domains.  (We know that the domains are microscopic because $\langle S_{max} \rangle /L \to 0$ in the thermodynamic limit.) This result is probably a consequence of the `weakness' of the uniform regime at $p=0$  discussed above.

\subsection{Field-free heterogeneous regime}

We turn now to the problem that motivated this paper, namely, the claim that the effect of a vanishingly small external field 
is to turn the field-free heterogeneous steady state  into a field-induced homogeneous state in the limit $L \to \infty$ \cite{Avella_06}.
This result would then be analogous to the homogenizing effect of an external noise that
changes traits at random with some small probability \cite{Klemm_03c}. Alas, we find no evidence of such mind-boggling effect neither in the one nor in the
two dimensional Axelrod's model. Actually, the original small-lattice simulations of Ref.\ \cite{Avella_06} do not seem to support the existence of
such phenomenon  either (see Sect. \ref{sec:2d}).

\begin{figure}[!ht]
  \begin{center}
\subfigure{\includegraphics[width=0.48\textwidth]{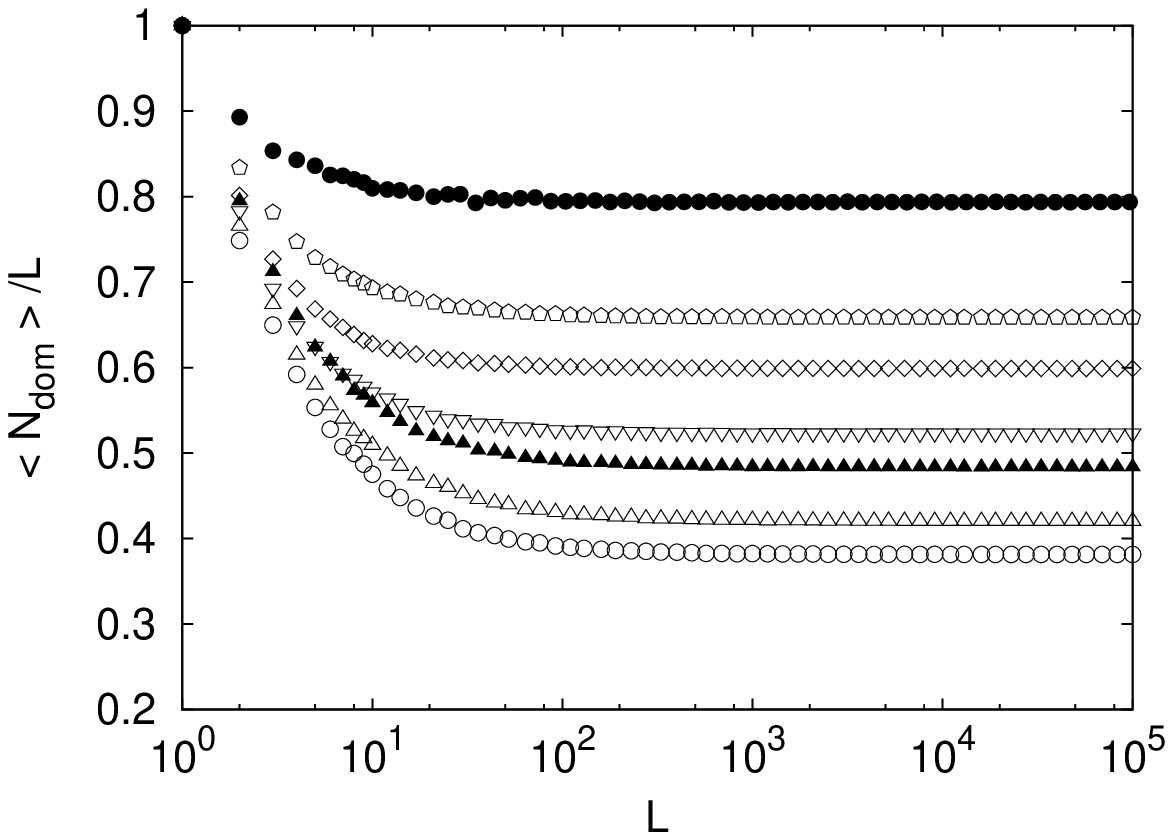}}
\subfigure{\includegraphics[width=0.48\textwidth]{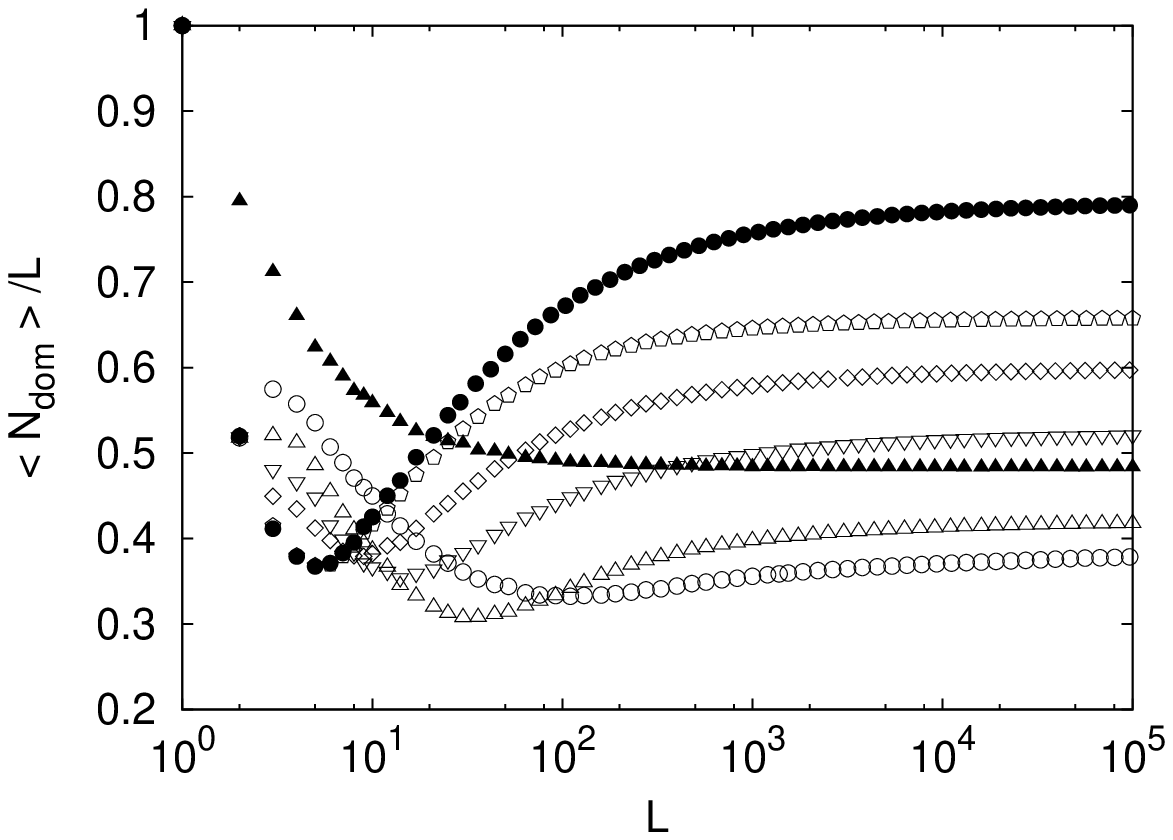}}
  \end{center}
\caption{Normalized number of domains in the one-dimensional model for the static (upper panel) and  consensus  (lower panel) media as function of the chain
size $L$.
The filled symbols are for $p = 1$ (circles)  and $p= 0$  (triangles) whereas  the  unfilled symbols are for (top to bottom at $L=10^5$)
$p= 0.99, 0.6, 0.4, 0.2$ and $0.01$.  The other parameters are $F = 5$ and $q = 10$.}
\label{fig:2}

\end{figure}

In Fig.\ \ref{fig:2} we show the average number of domains for both  static (upper panel) and consensus (lower panel)  media. We use filled symbols to display the results for 
the two extreme values of the field intensity ($p=0$ and $p=1$), and  unfilled symbols for the intermediary field strengths. This figure exhibits  many noticeable results.
The number of domains is maximum in the case only interactions with the field are permitted ($p=1$). Allowance of local interactions between neighboring agents results in
less heterogeneous configurations, as expected. The surprise is that, even for finite $L$, the average number of domains jumps to a lower value as $p$ departs
infinitesimally from unity (the same  phenomenon happens in the two-dimensional model \cite{Avella_06}). Although  $\langle N_{dom} \rangle $ exhibits a  smooth dependence on 
$p \in \left ( 0, 1 \right ) $  we observe another jump (now to a higher diversity value) at $p=0$. The fact that these jumps occur at any finite value of $L$ implies that the heterogeneous absorbing 
configurations are unstable to single-site changes. This conclusion is supported by the fact that the number of domains (and the diversity, as well) are reduced 
when the field or the local interactions are  turned on. Put differently, the number of domains decreases (discontinuously) from $p=0$ to $p >0$ as well as
from $p=1$ to $p < 1$.

We note that for $p=1$ the consensus media is fixed  with the traits reflecting the consensus values for the random initial configuration of the agents. However, for finite $L$ the results
differ somewhat significantly  from those of the static media (see Fig.\ \ref{fig:2}).  Why is that? The reason is that the traits of the static media are chosen randomly and independently of
the also random  initial traits of the  agents, whereas in the case of the consensus media the media traits are not independent from the agents' traits, being given by the majority rule.
This correlation has a strong effect for small chains  but becomes irrelevant for large $L$ since in this case the fraction of agents that share a given media trait is not much greater than $1/q$,
which is the expected  value of this fraction for the static media. In fact, the  correlation between the media and the agents in the  initial configuration explains the dips observed
in the lower panel of Fig.\ \ref{fig:2} for $p > 0$ since it  decreases the number of agents with antagonistic traits with respect to the media, resulting in less fragmented configurations 
than in the case the media traits are set randomly.  However, as pointed before, this effect disappears for large chains.

\begin{figure}[!ht]
  \begin{center}
\includegraphics[width=0.48\textwidth]{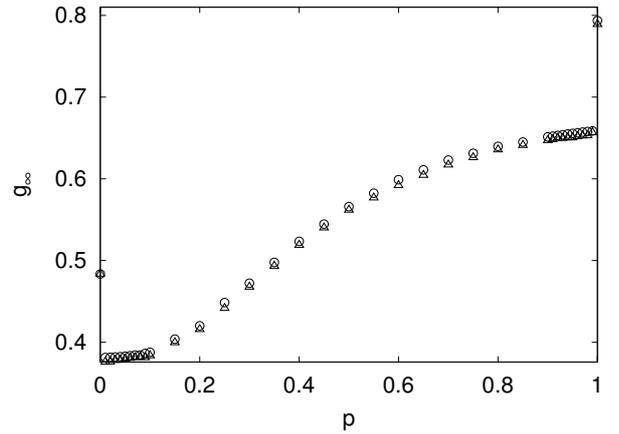}
  \end{center}
\caption{Extrapolation to $L \to \infty$ of the normalized number of domains in the one-dimensional model for the static (circles) and  consensus  (triangle) media as function of the 
field strength $p$. At $p=0$ we find $g_\infty = 0.483 \pm 0.001$  and at $p=1$ we find
$g_\infty = 0.793 \pm 0.001$  for both media types. The parameters are $F = 5$ and $q = 10$.}
\label{fig:3}
\end{figure}

The discontinuities at $p=0$ and $p=1$ are more easily visualized in   Fig.\ \ref{fig:3} where
 we present  the extrapolation to
infinite chain sizes, $g_\infty = \lim_{L \to \infty} \langle N_{dom} \rangle /L $, of the results exhibited in
Fig.\ \ref{fig:2}.  The results for both types of media  are practically indistinguishable.
For $p \to 0$ we find $g_\infty \to 0.38 \pm 0.01$ whereas  $g_\infty \to 0.66 \pm 0.01$ for $p \to 1$, which are valid for both media
types (see the legend of Fig.\ \ref{fig:3} for the values of $g_\infty$ at $p=0$ and $p=1$).   The important point here is that the data offer no 
evidence whatsoever that $g_\infty$ would vanish in the limit $p \to 0$ as claimed by Ref.\  \cite{Avella_06} for the two-dimensional model.
We should emphasize that the decrease on the number of domains induced by a vanishingly small field  is the expected outcome of the assay,
since the small field destabilizes some of the field-free domains,  but lacks the strength to create new field-induced domains.

In addition,  we find that the average size of the largest domain $\langle S_{max} \rangle $  grows as $\ln L$  for large $L$ (data not shown) whereas the size of a typical
domain is on the order of $1$, regardless of the media type and strength. These findings  are similar to those reported
 for  the majority-vote model \cite{Peres_10b}.

\section{External fields in the  two-dimensional model}\label{sec:2d}

We turn now to the  study of the  two-dimensional Axelrod's model,  which is considerably more computationally demanding than
 the analysis of the  one-dimensional version presented before. The effect of external fields on the field-free homogeneous regime
is well-understood by now \cite{Peres_11,Peres_10a}: the results are similar to those exhibited in  Fig.\ \ref{fig:1} except for
the relative size of the largest domain which in the limit $p \to 0$ tends to $1$ in the two-dimensional case \cite{Peres_11} and to $0$ in the one -dimensional 
version (lower panel of Fig.\ \ref{fig:1}). Hence we will consider  here only the effect of external fields on the field-free heterogeneous regime.
Moreover, we focus most our efforts on the parameter set $F=2$ and $q=8$ rather than on the
set   $F=5$ and $q=30$  of Ref.\ \cite{Avella_06}. In both cases the field-free ($p=0$) regime is heterogeneous, but simulations using our
parameter selection are much  faster of course. 

\begin{figure}[!ht]
  \begin{center}
\subfigure{\includegraphics[width=0.48\textwidth]{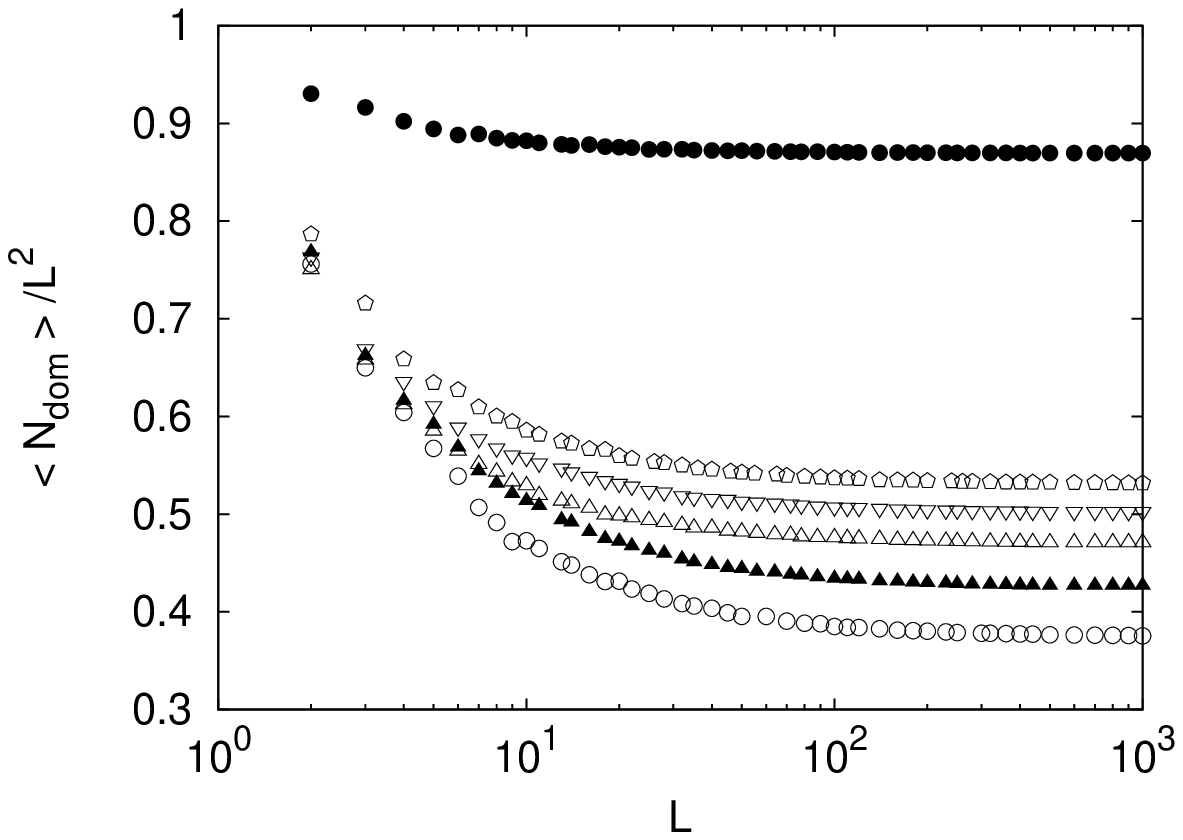}}
\subfigure{\includegraphics[width=0.48\textwidth]{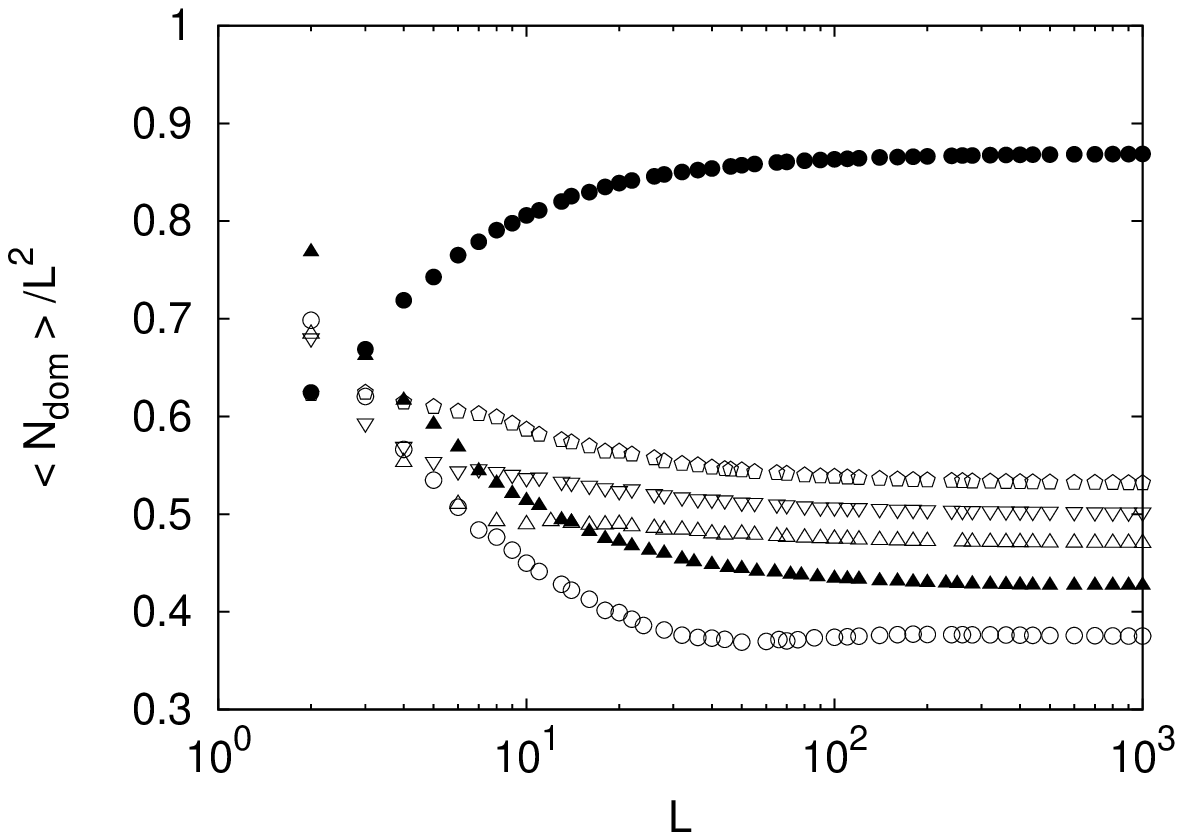}}
  \end{center}
\caption{Normalized number of domains in the two-dimensional model for the static (upper panel) and  consensus  (lower panel) media as function of the linear
size  $L$ of the square lattice.
The filled symbols are for $p = 1$ (circles)  and $p= 0$  (triangles) whereas  the  unfilled symbols are for (top to bottom at $L=10^3$)
$p= 0.99, 0.4, 0.2$ and $0.01$.  The other parameters are $F = 2$ and $q = 8$.}
\label{fig:4}

\end{figure}

\begin{figure}[!ht]
  \begin{center}
\includegraphics[width=0.48\textwidth]{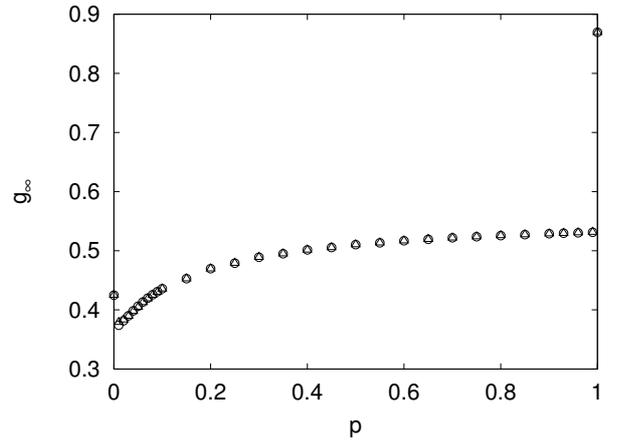}
  \end{center}
\caption{Extrapolation to $L \to \infty$ of the normalized number of domains in the two-dimensional model for the static (circles) and  consensus  (triangle) media as function of the 
field strength $p$.  The results are indistinguishable for the two media types. At $p=0$ we find $g_\infty = 0.425 \pm 0.001$  and at $p=1$ we find
$g_\infty = 0.869 \pm 0.001$. The parameters are $F = 2$ and $q = 8$.}
\label{fig:5}
\end{figure}

Accordingly,  in Fig.\ \ref{fig:4} we present the dependence of the average number of domains on the linear size of the square lattice. The results are
qualitatively similar to those obtained for the one-dimensional model and summarized in Fig.\ \ref{fig:2}. The  correlation between the consensus media and the
agents results  in less heterogeneous absorbing configurations in comparison with the configurations induced by the static media, but this difference becomes negligible
as the lattice size increases. This correlation effect is less dramatic than in the one-dimensional model because for the same  value of $L$  in the x-axis there are
many more agents ($L ^2$ to be precise)  in the two-dimensional lattice. The result of the extrapolation  to infinite lattice sizes using the
definition $g_\infty = \lim_{L \to \infty} \langle S_{dom} \rangle /L^2 $  are summarized in Fig.\ \ref{fig:5}.
For $p \to 0$ we find $g_\infty \to 0.36\pm 0.01$ whereas  $g_\infty \to 0.53 \pm 0.01$ for $p \to 1$. As before, we observe discontinuities at
$p=0$ and $p=1$ which take place even for finite $L$, and no evidence at all that the $g_\infty$ would tend to $0$ in the limit $p \to 0$ as claimed by 
Ref.\ \cite{Avella_06}.
We note, however, that extrapolation to $L \to \infty$ using lattices of linear size up to  $L=70$ as done in Ref.\ \cite{Avella_06} may result in a 
significant underestimate of $g_\infty$ since, especially for small $p$, one may be misled by the transient region where 
$ g_L \equiv \langle N_{dom} \rangle /L^2$  decreases abruptly with increasing $L$ (see Fig.\ \ref{fig:4}). 

Our simulations for the selection of parameters $F=5$ and $q=30$  used in Ref.\ \cite{Avella_06} led to  similar  conclusions. As already pointed out
the large relaxation times make an extensive analysis of this parameter set prohibitive, so we used a small number of samples (typically 100), which 
resulted in rather noisy data points.  For small lattice sizes ($L < 70$) our results fully  agree with those of \cite{Avella_06}. Actually, it should be
said that  those results do not support the claim that the $g_\infty$ vanishes in the limit $p \to 0$  for the static and the consensus media 
(see Fig. 7 of  Ref.\ \cite{Avella_06}). Perhaps this claim was motivated by an unwarranted generalization of the effect of a very distinct  type of field
as we discuss in the next section.

\section{Local  Field}\label{sec:disc}

Here we consider a time and spatially nonuniform field introduced in Ref.\ \cite{Avella_06}, referred to as local field (or media).  For a given target agent
the traits of this field reflect the consensus trait of its nearest neighbors.  So this field is a multi-state variant of the majority-vote rule 
\cite{Peres_10b,Krapivsky_92,Krapivsky_03,Parisi_03,Galam_05b,Lambiotte_08} 
and, in that sense, we are reluctant
to characterize it as a  media, much less as a mass media. In fact, its local nature sets  it apart from the static and consensus fields studied in the previous
sections, which are spatially uniform fields, and so we see no reason to assume the system will behave  similarly  under the effect of such 
diverse fields.

\begin{figure}[!ht]
  \begin{center}
\includegraphics[width=0.48\textwidth]{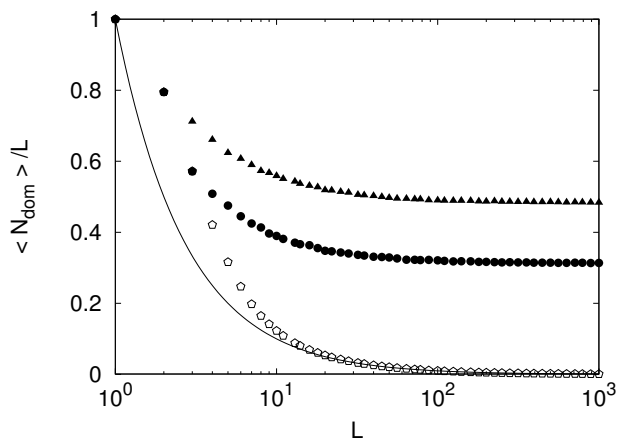}
  \end{center}
\caption{Normalized number of domains in the one-dimensional model for the local field as function of the chain
size $L$.
The filled symbols are for $p = 1$ (circles)  and $p= 0$  (triangles) whereas  the  unfilled symbols are for
$p= 0.99$. The solid  curve is $1/L$ and the other parameters are $F = 5$ and $q = 10$.}
\label{fig:6}
\end{figure}

The overly distinct effect of the local field on the steady state of Axelrod's model is illustrated in Fig.\ \ref{fig:6}  for the one-dimensional
variant.  Most interestingly, in this case the steady-state is less heterogeneous when the agents interact with the field only ($p=1$) than
when the field is off ($p=0$). This is a peculiarity  of the one-dimensional model which suits well to  illustrate the tendency to homogenization of the local field.
The results for all values of $p \in \left ( 0, 1 \right ) $ are indistinguishable from that exhibited in the figure for $p=0.99$, which shows the dominance of
homogeneous absorbing configurations  already for  small chains. In fact, for $L > 50$ the data points fall on the curve $1/L$ shown in the figure.
 So in the thermodynamic limit the steady state is spatially uniform except in the two extreme cases
$p=0$ and $p=1$.  A glance at  the results exhibited in Fig.\ \ref{fig:6} for the local field  and in Fig.\ \ref{fig:2} for the global fields reveals
the great disparity of the effect of these fields on the steady-state properties of Axelrod's model.

\begin{figure}[!ht]
  \begin{center}
\includegraphics[width=0.48\textwidth]{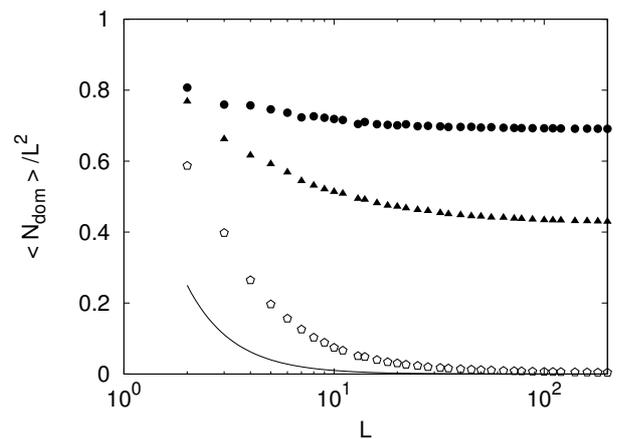}
  \end{center}
\caption{Normalized number of domains in the two-dimensional model for the local field as function of the linear
size $L$ of the square lattice.
The filled symbols are for $p = 1$ (circles)  and $p= 0$  (triangles) whereas  the  unfilled symbols are for
$p= 0.9$. The solid  curve is $1/L^2$ and the other parameters are $F = 2$ and $q = 8$.}
\label{fig:7}
\end{figure}

Figure \ref{fig:7} shows the results of the local field in the two-dimensional Axelrod model. These simulations are incredibly time consuming for $p \in \left ( 0, 1 \right )$ due to
the tendency to  homogenization of the agents' cultural traits and the constant need to update the local fields. Hence our simulations are restricted
to  $L < 200$ and $10^3$ samples only. As in the one-dimensional case, the simulation  data  in this
range of $p$ is practically indistinguishable within the numerical error and so we present only the results for $p=0.9$. However, 
the data do not fall on the curve $1/L^2$ (solid line in the figure), which would signal the  existence of a single domain. Rather, the scaled number of domains seems
to go to zero much slower than  $L^{-2}$ as the lattice size increases. This means that the average (non-normalized) number of domains grows with 
$L^\alpha$ with $\alpha < 2$, i.e., it  is non-extensive in presence of the local field.
 Of course, the small lattice sizes  as well as the reduced number of samples used in
this analysis does not allow us to make quantitative claims on the scaling laws for large $L$. 
We note that it was this difficulty to make inferences on the effect of the local field (even for a more manageable parameter set than that used in Ref.\
\cite{Avella_06}) that motivated our analysis of the one-dimensional model. Finally, we note that the absorbing configurations are more heterogeneous 
for $p=1$ than for $p=0$ in contrast to our findings for the one dimensional model.

\section{Conclusion}\label{sec:conc}

In our effort to clarify the effects of spatially uniform fields (interpreted as mass media) as well as of nonuniform fields  on 
Axelrod's  model  we have unveiled several interesting and unsuspected features of this well studied model of social influence.

First and foremost, we have debunked the claim that in the thermodynamic limit  the field-free heterogeneous regime becomes homogeneous in the limit of vanishingly small fields, i.e., $p \to 0$
regardless of the type of the field \cite{Avella_06}. More pointedly, we find that in the presence of a local field the steady state does indeed become homogeneous but for all $p \in \left ( 0, 1 \right )$.
This is hardly a surprise since the local field implements a majority-vote rule among the nearest neighbors of the target agent and so, when allied to the homogenizing interaction rule of
Axelrod's model,  it constitutes an insuperable force towards homogenization.  However, we find that the system remains heterogeneous in the presence of the spatially uniform fields which were 
originally introduced in Axelrod's model to study the effect of mass media on opinion formation \cite{Shibanai_01}.

Second, we find that the spatially uniform but time-varying consensus media introduced in \cite{Shibanai_01} and the static media yield the same results  in the thermodynamic limit.
It seems the reason they produce different outcomes for small lattices is the correlation between the consensus field and the agents in the  random initial configuration. This 
correlation becomes less pronounced for large lattices since even after application of the consensus rule there will be a rough balance between the values of the traits of a same  entry of
the feature vector.  From the statistical mechanics perspective, this is an important result because it allows us to make inferences about a more realistic media type using
the much easier to simulate static media.

Third, we find that the heterogeneous absorbing configurations at $p=0$ and $p=1$ are unstable. This instability is reflected by the discontinuities that
take place  at those values  even for finite lattices. Otherwise, the measures used to describe the steady state  -- average number of domains and
size of the largest domain -- are continuous functions
of $p$  in the range $p \in \left ( 0, 1 \right )$. 
We note that the field-free homogeneous configurations for finite $L$ are stable  in both one and two dimensions. However,
in the thermodynamic limit they become unstable in one dimension (lower panel of Fig.\ \ref{fig:1}) but remain stable in two dimensions \cite{Peres_11}.

Fourth, we find that the one-dimensional version of Axelrod's model \cite{Vilone_02} yields essentially the same results as the more popular two-dimensional
version. In particular,  our findings about the impact of the three field types on the field-free heterogeneous regime of  the two-dimensional model  were corroborated by
the one-dimensional model.  Although this model exhibits some peculiarities, which were properly highlighted in the text, it can serve as an exceptional 
guide to our understanding  of  features that are difficult to unveil in the two-dimensional version, such as the effect of  local fields.

The characterization of  the absorbing configurations of Axelrod's model  in the thermodynamic limit, regardless  of whether or not in the presence 
 of a field (media), remains a challenge to statistical mechanics. In fact, we do not know much about the location and the nature of the phase transition
 in the $\left ( q, F \right )$ space \cite{Castellano_00,Barbosa_09}, due mainly to the huge relaxation times the dynamics needs to  settle in a
 homogeneous configurations, which grows as $N^2$  where $N$ is the number of   lattice sites \cite{Vilone_02}. In that sense, the appearance of dubitable
 claims about the behavior of Axelrod's model in the thermodynamic limit should not be surprising. This situation is worsened by the presence of a field since
  lattices of intermediate sizes  exhibit  somewhat perversely a regime distinct from the thermodynamic one (see Fig.\ \ref{fig:1}). 

Nevertheless, we think  that  we have reached by now a good qualitative understanding of the effects of spatially   uniform (mass media) 
as well as nonuniform (local)  fields  on the statistical properties of  the
two-dimensional Axelrod model. In the field-free homogeneous regime case, the presence of a vanishingly small  field  (i.e., $ p \ll 1$) leads 
to the breaking of the single mono-cultural domain into a macroscopic domain (media) and a multitude of microscopic domains which
occupy a finite area of the lattice  \cite{Peres_11}. The scenario is different in the one-dimensional model where the field pulverizes the
giant domain into microscopic domains as shown in the lower panel of  Fig.\ \ref{fig:1}. Regarding the field-free heterogeneous regime case,
which was our main concern in this paper, the presence of a spatially uniform field does not produce cultural  homogenization, regardless of
the field strength $p$.   In contrast,  a local field  that implements the majority-vote rule among the neighbors of the target agent 
leads to a homogeneous steady state for $p < 1$.

\acknowledgments
This research was supported by The Southern Office of
Aerospace Research and Development (SOARD), Grant No.
FA9550-10-1-0006, and Conselho Nacional de Desenvolvimento
Cient\'{\i}fico e Tecnol\'ogico (CNPq).

\end{document}